# Process Algebra Based Tool Coordination Architectures in Raku and Go

*Bob Diertens*

B'ware Lab

*ABSTRACT*

This paper presents ongoing research in our project software engineering with process algebra. In this project we have developed among others a reimplementation of the simulator from the PSF Toolkit, a set of tools for the Process Specification formalism (PSF). This new simulator uses the ToolBus, a tool coordination architecture based on process algebra. We now developed new tool coordination architectures based on this ToolBus. We implement the primitives of the ToolBus in the programming languages Raku and Go. Both these languages have support for concurrency and communication between concurrent entities in the form of channels. We apply these tool coorination architectures on a small example. And we give implementations for the simulator in the PSF Toolkit based on the tool coordination architectures in Raku and Go.

*Keywords:* process algebra, software engineering, software architecture, tool coordination architecture,

## 1. Introduction

In our project *Software Engineering with Process Algebra* [1] we applied process algebra on several levels of abstractions in the design. We use the Process Specification Formalism (PSF) [7] as our process algebra. PSF is accompanied by a Toolkit that contains, among others, a compiler and a simulator. As target tool coordination architecture for the implementation of some of the designs we used the ToolBus. The ToolBus acts as a simulator for process algebra based scripts. In [5] we reengineered the simulator from the PSF Toolkit into an ToolBus application.

The language for writing the ToolBus script can be seen as a special purpose process algebra language, solely for tool coordination specification. Instead of using the ToolBus acting as a simulator for the coordination of the tools, we like to be able to implement tool coordination using features such as concurrency that are provided by modern programming languages. That gives us more flexibilty in implementation and communication with the tools. For instance, a tool could be implemented as a routine that runs in concurrence with the tool coordination. In this paper, we describe our work on implementations of a tool coordination architecture using the programming languages Raku and Go.

Raku [2] is a programming language that originates from Perl [9]. Perl 6, as a followup from Perl 5, differed so much from its predecessor that it has been renamed into Raku in 2019 and became a programming language on its own. Go [6] [3] is a programming language designed at Google in 2007, that was first announced publicly in 2009. The reason for using Raku and Go for our implementation of a tool coordination architecture is that they both support concurrency and use channels for communication between concurrent entities.

In section 2 we give a brief description of the ToolBus with an example of its use. The implementation of the tool coordination architectures in Raku and Go is given in section 3, including the application to the example given in the section on the ToolBus. In section 4 we describe the application of the tool coordination architectures on the simulator of the PSF Toolkit. And in section 5 we give some comcluding remarks.

## 2. ToolBus

The ToolBus coordination architecture [4] is a software application architecture developed at CWI (Amsterdam) and the University of Amsterdam. It utilizes a scripting language based on process algebra to describe the communication between software tools. A ToolBus script describes a number of processes that can communicate with each other and with various tools existing outside the ToolBus. The role of the ToolBus when executing the script is to coordinate the various tools in order to perform some complex task. A language-dependent adapter that translates between the internal ToolBus data format and the data format used by the individual tools makes it possible to write every tool in the language best suited for the task(s) it has to perform.

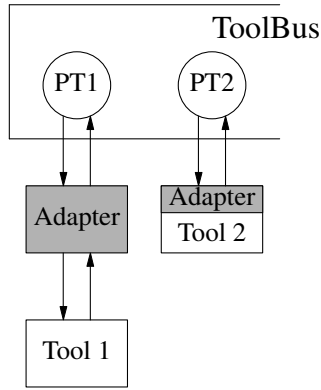

**Figure 1.** Model of tool and ToolBus interconnection

We give a minimal description of the ToolBus, just enough for our purposes. We refer to the user guide distributed with the ToolBus software package for a complete description. In Figure 1 two possible ways of connecting tools to the ToolBus are displayed. One way is to use a separate adapter and the other to have a built-in adapter. Communications between ToolBus processes can be done using the primitives `snd-msg` and `rec-msg`. A ToolBus process can communicate with a tool using the primitives `snd-do` and `snd-eval`. With the latter the tool has to send back a value which the ToolBus process can receive with the primitive `rec-value`. A tool can send an event to a ToolBus process that is to be received with the primitive `rec-event`, to be acknowledged by the ToolBus process using the primitive `snd-ack-event`. An overview of possible communications inside the ToolBus and with the tools is given in Table 1, here `<function>` represents the function to be called by the adapter of the tool.

**Table 1.** ToolBus communications

| ToolBus process | ToolBus process |
|---|---|
| `snd-msg(`*Term*`, ...)` | `rec-msg(`*Term*`, ...)` |
| **ToolBus process** | **Tool** |
| `snd-do(`*ToolID*`, <function>(`*arg*`, ...))` | `<function>` |
| `snd-eval(`*ToolID*`, <function>(`*arg*`, ...))` | `<function>` |
| `snd-ack-event(`*ToolID*`, `*Term*`)` | `<rec-ack-event>` |
| **Tool** | **ToolBus process** |
| `snd-value(`*Term*`)` | `rec-value(`*ToolID*`, `*Term*`)` |
| `snd-event(`*Term*`, ...)` | `rec-event(`*ToolID*`, `*Term*`, ...)` |

*2.1 Example*

As an example of the use of the ToolBus, we specify an application carried out in the form as shown in Figure 1. In this example, Tool1 can either send a 'message' to Tool2 and then wait for an acknowledgement from Tool2, or it can send a 'quit' after which the application will shutdown.

The implementation consists of three Tcl/Tk[1] [8] programs (Tool1, its adapter, and Tool2), and a ToolBus script. A screendump of this application at work together with the viewer of the ToolBus is shown in Figure 2. With the viewer it is possible to step through the execution of the ToolBus script and view the variables of the individual processes inside the ToolBus.

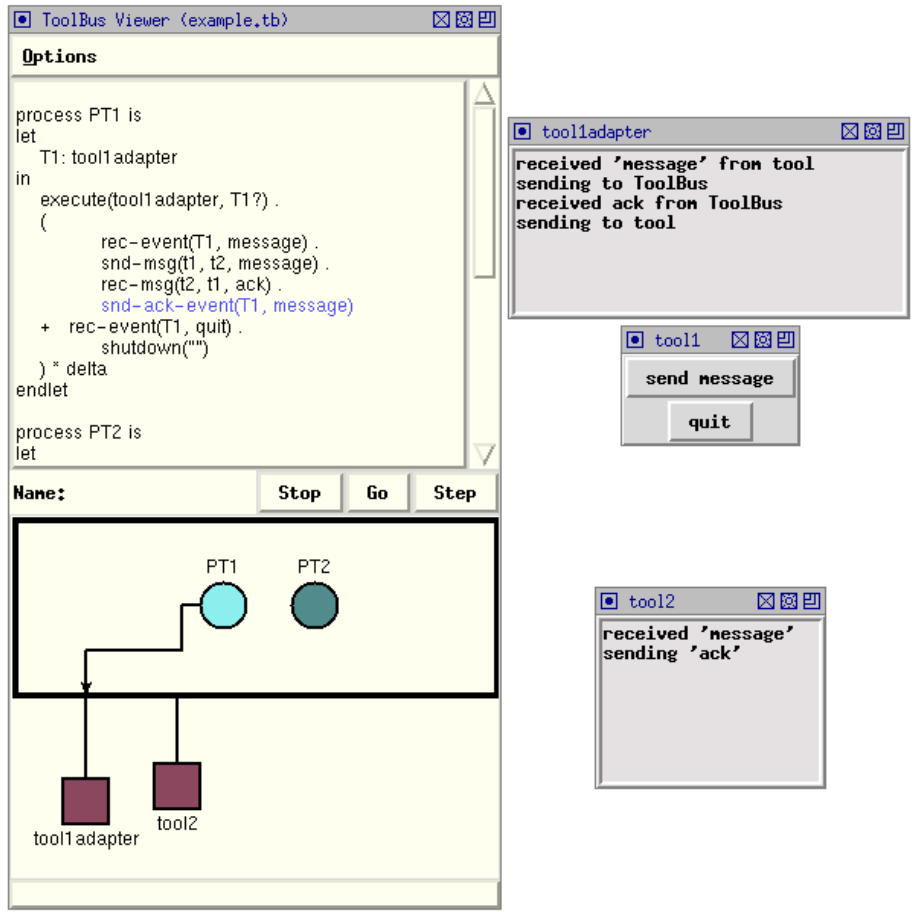

**Figure 2.** Screendump of the example as ToolBus application with viewer

The ToolBus script is shown below. The `execute` actions in the ToolBus script correspond to the starting of the adapter for Tool1 and the starting of Tool2 in parallel with the processes `PT1` and `PT2` respectively. The variables `T1` and `T2` are used as identifiers for Tool1 and Tool2, and the terms `t1` and `t2` serve as identifiers for the ToolBus processes `PT1` and `PT2`.

```
        process PT1 is                                                                toolbus
     let
        T1: tool1adapter
     in
        execute(tool1adapter, T1?) .
        (
           rec-event(T1, message) .
           snd-msg(t1, t2, message) .
           rec-msg(t2, t1, ack) .
           snd-ack-event(T1, message)
        +  rec-event(T1, quit) .
           shutdown("")
        ) * delta
```

---

1.  Tcl/Tk is combination of the tool command language Tcl and the Tk GUI toolkit extension package.

```
      endlet

      process PT2 is
      let
         T2: tool2
      in
         execute(tool2, T2?) .
         (
            rec-msg(t1, t2, message) .
            snd-eval(T2, eval(message)) .
            rec-value(T2, value(ack)) .
            snd-msg(t2, t1, ack)
         ) * delta
      endlet

      tool tool1adapter is {
         command = "wish-adapter -script tool1adapter.tcl" }
      tool tool2 is { command = "wish-adapter -script tool2.tcl" }

      toolbus(PT1, PT2)
```

Following the description of the ToolBus processes is the description of how to execute the tools by the execute actions. The last line of the ToolBus script starts the processes `PT1` and `PT2` in parallel.

### 3. Tool Coordination Architecture Primitives

We implement the primitives of the ToolBus, so that we can build an application simular to a ToolBus script. The primitives are processes, communication between the processes, tool connections, communcations between these connections and the processes, and operations for the coordination architecture.

However, an implementation based on these primitives differs in operation with an implementation that uses the ToolBus. We list these differences below.

- The Toolbus uses sockets (Unix domain sockets and network sockets) to commuicate with the tools. We communicate with the tools over pipelines, but other ways can always be added.

- The ToolBus simulates a script in a single process. We run processes concurrently.

- The ToolBus takes care of the communication with the tools. We let each a process take care of the communtions with the tool which the process starts up itselfs.

- The ToolBus arranges for all communications between the processes to take place. The processes have to take care of the communications they are involved in to happen.

- The adapters provided by the ToolBus for the tools take care of calling the right function of the tools. We let the tools themselves to take care of this.

- The communications between the processes and between the tools and processes are split up by their function within the ToolBus. This the tools must do themselves.

*3.1 Implementation*

For the implementation of the primitves in Raku and Go, we use a module/package that we call `BTCA` (Bob's Tool Coordination Architecture). The source code for the Raku package can be found in appendix A.1 and for the Go module in appendix A.2.

*3.1.1 Processes amd communication between processes*

In both Raku and Go we use functions for implementing the processes. These function are supposed to be run in concurrence with eachother and take care of handling the communication between them. In Raku concurrent function execution can be done with the command `start` and in Go with the command `go`.

Raku and Go both have channels for communications between concurrent functions. In Go we use unbuffered channels for the communication between the processes. The channels in Raku are always buffered unfortunately. We mimick unbuffered channels with two buffered channels, one for sending the message and one for receiving an acknowledgement message. By always waiting for an acknowledgement on sending a message and by directly sending an acknowledgement after receiving a message, we effectively have the behaviour of unbuffered channels. We define a new class `Channel` with the methods `snd-msg` and `rec-msg` for the sending and receiving of messages. The interface for these channels can be found in Table 2.

**Table 2.** Channel interface

|  | Raku | Go |
|---|---|---|
| **instance** | `$ch = BTCA::Channel.new(<id>)` | `ch = make(chan string)` |
| **send** | `$ch.snd-msg(<message>)` | `ch <- <message>` |
| **receive** | `$s = $ch.rec-msg()` | `s = <- ch` |

The `<id>` in the creation of a new channel is there for debugging purposes only. The Raku version of receive has an option to make it an blocking operation for when it is not used in a communication event handler construction (see section 3.1.3). It can be used as in `$s = $ch.rec-msg(:block)`.

*3.1.2 Tools and communication with processes*

We provide a standard way for running tools, in the form of a class with methods for creation, starting, receiving, sending, and stopping. For the communication between the tools and proceses we use channels. The interface can be found in Table 3.

**Table 3.** Tool interface

|  | Raku | Go |
|---|---|---|
| **instance** | `$tool = BTCA::Tool.new(<id>, <command>, <args>)` | `tool = BTCA.NewTool(<id>, <command>, <args>)` |
| **start** | `$tool.Start()` | `tool.Start()` |
| **send** | `$tool.Send(<message>)` | `tool.Send <- <message>` |
| **receive** | `$s = $tool.Receive()` | `s = <- tool.Receive` |
| **stop** | `$tool.Kill()` | `tool.Kill()` |

The `<id>` in the creation of new tool instances is there for debugging purposes only. As with the receive operation from the channel interface, the receive operation of the tool interface has also an option to make it a blocking operation.

We have defined expressions for extracting the argument of the primitives out of the received strings from the tools. They are conveniently named after the counterparts of the primitives in the communication.

**Table 4.** Expressions for argument exraction

|  | Raku |
|---|---|
| **snd-event** | `$s ~~ %BTCA::expr{'rec-event'}` |
| **snd-value** | `$s ~~ %BTCA::expr{'rec-value'}` |
|  | Go |
| **snd-event** | `m := BTCA.Expr["rec-event"].FindStringSubmatch(s)` |
| **snd-value** | `m := BTCA.Expr["rec-value"].FindStringSubmatch(s)` |

For Raku, the result of the matched argument can be found in `$0`, and for Go, the result can be found in `m[1]`.

*3.1.3 Communication event handlers*

The ToolBus handles all communications for the processes and the tools. In Raku and Go each process has to handle its own communication. Naturally, these languages provide support for handling these communication events. This handling of communication events between processes and between processes and tools typically have the following form in Raku.

```raku
react {
    whenever $tool.Receive() {
       ...
    }
    whenever $ch.rec-msg() {
       ...
    }
    ...
}
```

The handler can be stopped with a `done()` statement.

In Go such a communication event handler can have to following form.

```go
run := true
for run {
   select {
   case a := tool.Receive:
       ...
   case a := <- ch:
       ...
   }
   ...
}
```

This handler can be stopped with setting the variable `run` to `false`. Another way have jumping out of the `for` loop is also possible, for instance with the use of a labelled `break` statement.

*3.1.4 TCA Operations*

The tool coordination architecture modules have operation for starting and stopping the tool coordination, as well as setting on/off a debugging mode.

**Table 5.** Coordination Architecture Operations

|  | Raku | Go |
|---|---|---|
| **start** | `BTCA::Run(<function> ...)` | `BTCA.Run(<function> ...)` |
| **stop** | `BTCA::Shutdown()` | `BTCA.Shutdown()` |
| **debug** | `BTCA::Debug(<boolean>)` | `BTCA.Debug(<boolean>)` |

The `Run` operation starts the concurrent execution of all the functions given as argument. The `Debug` operation turns on or off the debugging mode. When it is turned on, information about the communication between the tools and the proceses is shown, and in the case of Raku also the communication between the processes. In Go we use for the communication between the processes the channels as provided by Go, and do not have the possibility to let them generate some debugging information.

*3.2 Example*

We give implementations for the example from section 2 in both Raku and Go using the tool coordination architecture modules.

```raku
#!/usr/bin/raku

use lib '/path_to_lib';
use BTCA;

my $toolpath = '/path_to_tools';

BTCA::debug(False);

my $chan12 = BTCA::Channel.new('12');
my $chan21 = BTCA::Channel.new('21');
```

```perl
sub comp1() {
    my $comp = BTCA::Tool.new('comp1', '/usr/bin/wish',
        "$toolpath/tool1adapter.tcl");
    $comp.Start();
    react {
        whenever $comp.Receive() {
            $_ ~~ %BTCA::expr{'rec-event'};
            my $a = $0;
            if ($a eq 'quit') {
                $comp.Send('quit');
                BTCA::Shutdown('x');
            } else {
                $chan12.snd-msg($a);
                $chan21.rec-msg(:block);
                $comp.Send("snd-ack-event($a)");
            }
        }
    }
}
sub comp2() {
    my $r;

    my $comp = BTCA::Tool.new('comp2', '/usr/bin/wish',
        "$toolpath/tool2.tcl");
    $comp.Start();
    react {
        whenever $chan12.rec-msg() {
            $r = $_;
            if /^quit$/ {
                $comp.Kill();
                done();
            } else {
                $comp.Send("snd-eval($r)");
                $r = $comp.Receive(:block);
                $r ~~ %BTCA::expr{'rec-value'};
                my $a = $0;
                $chan21.snd-msg($a);
            }
        }
    }
}

BTCA::Run(&comp2, &comp1);
```

```go
package main

import (
    "/path_to_lib/BTCA"
)

var toolpath = "/path_to_tools"

var ch = map[string] chan string {
    "12" : make(chan string),
    "21" : make(chan string),
}

func comp1() {
    comp := BTCA.NewTool("comp1", "/usr/bin/wish",
        toolpath + "/tool1adapter.tcl")
    comp.Start()
    run := true
    for run {
        select {
        case a := <- comp.Receive:
            m := BTCA.Expr["rec-event"].FindStringSubmatch(a)
            if m[1] == "quit" {
```

```
                comp.Send <- "quit"
                BTCA.Shutdown()
            } else {
                ch["12"] <- m[1]
                a = <- ch["21"]
                comp.Send <- "snd-ack-event(ack)"
            }
        }
    }
}
func comp2() {
    comp := BTCA.NewTool("comp2", "/usr/bin/wish", toolpath + "/tool2.tcl")
    comp.Start()
    run := true
    for run {
        select {
        case a := <- ch["12"]:
            if a == "quit" {
                comp.Kill()
                run = false
                break;
            } else {
                comp.Send <- "snd-eval(" + a + ")"
                a := <- comp.Receive
                m := BTCA.Expr["rec-value"].FindStringSubmatch(a)
                if m != nil {
                    ch["21"] <- "ack"
                }
            }
        }
    }
}
func main() {
    BTCA.Run(comp1, comp2)
}
```

## 4. Tool Coordination Architecture based PSF Simulator

We have implemented the tool coordination architecture for the PSF simulator in both Raku and Go. We followed thereby the ToolBus implementation of the PSF simulator with some adaptions for the differences in implementation of the primitives, as mentioned in section 3. We also had to make some changes to the adapters of the tools in accordance with these differences.

The implementations of the tool coordination architecture for the PSF simulator can be found in appendix B.

## 5. Conclusions

We have implemented primitives for tool coordination architectures in both Raku and Go, where we followed the implentation of the ToolBus. Based on these implementations we build new implementation for the simulator of the PSF Toolkit in Raku and Go. The ToolBus implementation of the simulator was based on design specification in process algebra (see [5]). The new implementation for the simulator are still conform these process algebra specification.

The tool coordination architectures that we implemented are limited to our needs. The implementation are however more flexible than the ToolBus. They allow for instance implementation of a tool as a concurrent function, instead of as a seperate tool. And they can be extended with other features. We see no problems in extending the implementation with things as tool communication over sockets, flexibel creation and destroying of processes (with or without tools). Also, extending the debugging possibilites to something as a viewer similar to the one provided by the ToolBus is possible.

## A. BTCA modules

*A.1 Raku*

```raku
unit module BTCA;                                                          raku

my $DEBUG = False;

our sub debug($d) {
   $DEBUG = $d;
}
class Channel {
   has $.id;
   has @.channel;

   method new($id) {
      my @channel;
      push @channel, CORE::Channel.new;
      push @channel, CORE::Channel.new;
      self.bless(:$id, :@channel);
   }

   method snd-msg($s) {
      my $ack;
      if $DEBUG { say "TCA chan snd ", $!id, " : $s"; }
      @!channel[0].send($s);
      $ack = @!channel[1].receive;
      if $DEBUG { say "TCA chan $ack "; }
   }

   method rec-msg (:$block = False) {
      if $block {
         my $r;
         $r = @!channel[0].receive;
         self!ackout($r);
         return $r;
      } else {
         supply whenever @!channel[0] {
            self!ackout($_);
            .emit;    # equals self.emit and $_.emit, but emit of what?
         }
      }
   }

   method !ackout($s) {
      @!channel[1].send("$!id sync: $s");
   }

}

my @toollist = ();

class Tool is Proc::Async {
   has $.id;
   has $.tool;
   has $.channel;

   method new($id, $path, *@args) {
      my $t = Proc::Async.new(:w, $path, @args);
      my $c = $t.stdout.lines.Channel();
      self.bless(:$id, tool => $t, channel => $c);
   }

   method Start($me:) {
      push @toollist, $me;
      $!tool.start;
   }

   method Receive(:$block = False) {
```

```
        if $block {
            my $r;
            $r = $!channel.receive;
            return $r;
        } else {
            supply whenever $!channel {
                my $s = chomp $_;
                if $DEBUG { say "TCA " ~ $!id ~ " receive: $s"; }
                emit($s);
            }
        }
    }

    method Send($s) {
        $!tool.say($s);
        if $DEBUG { say "TCA " ~ $!id ~ " send: $s"; }
    }

    method Kill() {
        if $DEBUG { say "TCA tool $!id killed"; }
        $!tool.kill();
    }
}

our %expr = Map.new(
    'rec-event', /^snd\-event\((.+)\)$/,
    'rec-value', /^snd\-value\((.+)\)$/,
);

my @processlist = ();

our sub Run(*@funcs) {
    my $p;
    my $f;

    for @funcs -> $f {
        $p = start $f();
        push @processlist, $p;
    }
    await @processlist;
}

our sub Shutdown($m) {
    my $t;

    for @toollist -> $t {
        # $t.snd-terminate($m);
        # or
        $t.Kill();
    }
    exit(0);
}
```

## A.2 Go

```go
package BTCA

import (
    "bufio"
    "os/exec"
    "os"
    "strings"
    "fmt"
    "container/list"
    "sync"
    "regexp"
)
```

```go
var DEBUG = false

func Debug(d bool) {
   DEBUG = d
}

var Expr = map[string] *regexp.Regexp {
   "rec-event" : regexp.MustCompile(`^snd-event\((.+)\)$`),
   "rec-value" : regexp.MustCompile(`^snd-value\((.+)\)$`),
}

var toollist = list.New()

type Tool struct {
   id string
   cmd *exec.Cmd
   reader *bufio.Reader
   writer *bufio.Writer
   Send chan string
   Receive chan string
}

func NewTool(id string, c string, a ...string) Tool {
   var t Tool

   t.id = id
   t.cmd = exec.Command(c, a...)
   stdout, _ := t.cmd.StdoutPipe()
   stdin, _ := t.cmd.StdinPipe()
   t.reader = bufio.NewReader(stdout)
   t.writer = bufio.NewWriter(stdin)
   t.Send = make(chan string)
   t.Receive = make(chan string)
   toollist.PushFront(t);
   return t
}

func (t Tool) Start() {
   t.cmd.Start()
   go t.sendtool()
   go t.receivetool()
}

func (t Tool) sendtool() {

   for {
      select {
      case s := <- t.Send:
         t.writer.WriteString(s + "0)
         t.writer.Flush()
         if DEBUG { fmt.Printf("TCA %s send: %s0, t.id, s) }
      }
   }
}

func (t Tool) receivetool() {

   for {
      txt, err := t.reader.ReadString('0)
      if DEBUG { fmt.Printf("TCA %s receive: %s", t.id, txt) }
      if err != nil {
         return
      } else {
         txt, _ = strings.CutSuffix(txt, "0)
         t.Receive <- txt
      }
   }
}

func (t Tool) Kill() {
   t.cmd.Process.Kill()
   if DEBUG { fmt.Printf("TCA tool %s killed0, t.id) }
```

```
}
func Run(procs ...func()) {

    var wg sync.WaitGroup

    for _, p := range procs {
        wg.Add(1)
        go p()
    }
    wg.Wait()
}
func Shutdown() {
    for e := toollist.Front(); e != nil; e = e.Next() {
        tool := Tool(e.Value.(Tool))
        tool.Kill()
    }
    os.Exit(0)
}
```

## B. PSF Simulator

The implementations of the tool coordination for the simulator from the PSF Toolkit given here as ToolBus script and in the programming languages Raku and Go all come with a script in Perl to start them up. The script takes care of collecting the options and further arguments from the command line, and communicates them to the different tools using environment variables.

*B.1 ToolBus*

```
                                                                                    toolbus
        process PGUI is
        let GUI : gui,
           S1: str,
           S2: str,
           S3: str,
           S4: str,
           S5: str,
           S6: str
        in
           execute(gui, GUI?) .
           rec-event(GUI, window(S1?, S2?, S3?, S4?, S5?, S6?)) .
           snd-ack-event(GUI, window(S1, S2, S3, S4, S5, S6)) .
           snd-msg(gui, function, window(S1)) .
           snd-msg(gui, startprocess, window(S2)) .
           snd-msg(gui, tracectrl, window(S3)) .
           snd-msg(gui, breakctrl, window(S4)) .
           snd-msg(gui, display, window(S5)) .
           snd-msg(gui, actionchooser, window(S6))
        endlet

        process PKERNEL is
        let KERNEL : kernel,
           S : str,
           A : str,
           T : str,
           I : int,
           J : int,
           Wait : bool
        in
           execute(kernel, KERNEL?) .
           snd-eval(KERNEL, get-action-info) .
           rec-value(KERNEL, action-info(S?)) .
           snd-msg(kernel, tracectrl, action-info(S)) .
           snd-msg(kernel, breakctrl, action-info(S)) .
           snd-eval(KERNEL, get-process-list) .
           rec-value(KERNEL, process-list(S?)) .
           snd-msg(kernel, startprocess, process-list(S)) .
           Wait := true .
           (
              if not(Wait) then
                 snd-eval(KERNEL, compute-choose-list) .
                 (
                    rec-value(KERNEL, action-choose-list(S?)) .
                    snd-msg(kernel, actionchooser, action-choose-list(S))
                 +  rec-value(KERNEL, halt(T?)) .
                    snd-msg(kernel, actionchooser, halt(T)) .
                    snd-msg(kernel, display, halt(T))
                 ) .
                 Wait := true
              else
                 rec-msg(actionchooser, kernel, action(info(I?, J?, S?, A?))) .
                 snd-do(KERNEL, action(I, J, A)) .
                 Wait := false
              +  rec-msg(startprocess, kernel, reset) .
                 snd-do(KERNEL, myreset) .
                 snd-msg(kernel, display, reset) .
                 snd-msg(kernel, actionchooser, reset) .
                 Wait := false
              +  rec-msg(function, kernel, quit) .
                 snd-eval(KERNEL, quit) .
```

```
            rec-value(KERNEL, quit) .
            snd-msg(kernel, display, quit) .
            rec-msg(display, kernel, quit) .
            shutdown("")
      +   rec-msg(function, kernel, process-status) .
            snd-eval(KERNEL, process-status) .
            rec-value(KERNEL, process-status(S?)) .
            snd-msg(kernel, display, process-status(S))
      +   rec-msg(startprocess, kernel, start(I?, S?)) .
            snd-do(KERNEL, start(I, S)) .
            snd-msg(kernel, display, start(S)) .
            snd-msg(kernel, actionchooser, reset) .
            Wait := false
      +   rec-msg(actionchooser, kernel, save(I?)) .
            snd-do(KERNEL, save(I))
      +   rec-msg(actionchooser, kernel, goto(I?)) .
            snd-do(KERNEL, mygoto(I)) .
            Wait := false
         fi
      ) * delta
endlet

process PPROCESS is
let PROCESS : startprocess,
    S : str,
    I : int
in
    rec-msg(gui, startprocess, window(S?)) .
    execute(startprocess(S), PROCESS?) .
    rec-msg(kernel, startprocess, process-list(S?)) .
    snd-do(PROCESS, process-list(S)) .
    (
       rec-event(PROCESS, start(I?, S?)) .
       snd-ack-event(PROCESS, start(I, S)) .
       snd-msg(startprocess, kernel, start(I, S))
     + rec-event(PROCESS, reset) .
       snd-ack-event(PROCESS, reset) .
       snd-msg(startprocess, kernel, reset)
    ) * delta
endlet

process PACTIONCHOOSER is
let ACTIONCHOOSER : actionchooser,
    S : str,
    A : term,
    T : term,
    I : int,
    Random : bool,
    Choosing : bool
in
    rec-msg(gui, actionchooser, window(S?)) .
    execute(actionchooser(S), ACTIONCHOOSER?) .
    Random := false .
    Choosing := false .
    (
       rec-msg(kernel, actionchooser, action-choose-list(S?)) .
       if Random then
          snd-msg(actionchooser, breakctrl, action-choose-list(S)) .
          (
             rec-msg(breakctrl, actionchooser, break) .
             Random :=false .
             snd-do(ACTIONCHOOSER, random-off)
           + rec-msg(breakctrl, actionchooser, action-choose-list(S?))
          )
       else
          tau
       fi .
       snd-do(ACTIONCHOOSER, action-choose-list(S)) .
       Choosing := true
     + rec-msg(kernel, actionchooser, halt(S?)) .
       snd-do(ACTIONCHOOSER, random-off) .
       Random := false .
```

```
         snd-do(ACTIONCHOOSER, halt) .
         Choosing := true
   +  if Choosing then
         rec-event(ACTIONCHOOSER, action(A?)) .
         snd-ack-event(ACTIONCHOOSER, action(A)) .
         Choosing := false .
         (
            snd-msg(actionchooser, kernel, action(A)) .
            if Random then
               snd-msg(actionchooser, breakctrl, action(A)) .
               (
                  rec-msg(breakctrl, actionchooser, break) .
                  Random :=false .
                  snd-do(ACTIONCHOOSER, random-off)
               +  rec-msg(breakctrl, actionchooser, nobreak) .
                  snd-msg(actionchooser, tracectrl, action(A)) .
                  rec-msg(tracectrl, actionchooser, done)
               )
            else
               snd-msg(actionchooser, tracectrl, action(A)) .
               rec-msg(tracectrl, actionchooser, done)
            fi
         +  rec-msg(kernel, actionchooser, reset) .
            snd-do(ACTIONCHOOSER, reset) .
            Choosing := false
         )
      +  rec-event(ACTIONCHOOSER, save(I?)) .
         snd-ack-event(ACTIONCHOOSER, save(I)) .
         (
            snd-msg(actionchooser, kernel, save(I))
         +  rec-msg(kernel, actionchooser, reset) .
            snd-do(ACTIONCHOOSER, reset) .
            Choosing := false
         )
      +  if not(Random) then
            rec-event(ACTIONCHOOSER, goto(I?)) .
            snd-ack-event(ACTIONCHOOSER, goto(I)) .
            (
               snd-msg(actionchooser, kernel, goto(I)) .
               Choosing := false
            +  rec-msg(kernel, actionchooser, reset) .
               snd-do(ACTIONCHOOSER, reset) .
               Choosing := false
            )
         fi
      fi
   +  rec-msg(kernel, actionchooser, reset) .
      snd-do(ACTIONCHOOSER, reset)
   +  rec-event(ACTIONCHOOSER, random(T?)) .
      snd-ack-event(ACTIONCHOOSER, random(T)) .
      Random := equal(T, on)
   ) * delta
endlet

process PFUNCTION is
let FUNCTION : function,
    S : str,
    T : term
in
   rec-msg(gui, function, window(S?)) .
   execute(function(S), FUNCTION?) .
   (
      rec-event(FUNCTION, quit) .
      snd-ack-event(FUNCTION, quit) .
      snd-msg(function, kernel, quit)
   +  rec-event(FUNCTION, process-status) .
      snd-ack-event(FUNCTION, process-status) .
      snd-msg(function, kernel, process-status)
   ) * delta
endlet

process PDISPLAY is
```

```
   let DISPLAY : display,
      T : term,
      S : str
   in
      rec-msg(gui, display, window(S?)) .
      execute(display(S), DISPLAY?) .
      (
         rec-msg(kernel, display, halt(S?)) .
         snd-do(DISPLAY, halt(S))
      +  rec-msg(kernel, display, start(S?)) .
         snd-do(DISPLAY, start(S))
      +  rec-msg(kernel, display, reset) .
         snd-do(DISPLAY, reset)
      +  rec-msg(kernel, display, process-status(S?)) .
         snd-eval(DISPLAY, process-status(S)) .
         rec-value(DISPLAY, done)
      +  rec-msg(tracectrl, display, trace(T?)) .
         snd-eval(DISPLAY, trace(T)) .
         rec-value(DISPLAY, done)
      +  rec-msg(breakctrl, display, break(T?)) .
         snd-do(DISPLAY, break-action(T))
      +  rec-msg(breakctrl, display, break) .
         snd-do(DISPLAY, break)
      +  rec-msg(kernel, display, quit) .
         snd-eval(DISPLAY, quit) .
         rec-value(DISPLAY, quit) .
         snd-msg(display, kernel, quit)
      ) * delta
   endlet

   process PTRACECTRL is
   let TRACECTRL : tracectrl,
      T : term,
      S : str
   in
      rec-msg(gui, tracectrl, window(S?)) .
      execute(tracectrl(S), TRACECTRL?) .
      rec-msg(kernel, tracectrl, action-info(S?)) .
      snd-do(TRACECTRL, action-info(S)) .
      (
         rec-msg(actionchooser, tracectrl, action(T?)) .
         snd-eval(TRACECTRL, action(T)) .
         (
            rec-value(TRACECTRL, trace) .
            snd-msg(tracectrl, display, trace(T))
         +  rec-value(TRACECTRL, notrace)
         ) .
         snd-msg(tracectrl, actionchooser, done)
      ) * delta
   endlet

   process PBREAKCTRL is
   let BREAKCTRL : breakctrl,
      T : term,
      S : str
   in
      rec-msg(gui, breakctrl, window(S?)) .
      execute(breakctrl(S), BREAKCTRL?) .
      rec-msg(kernel, breakctrl, action-info(S?)) .
      snd-do(BREAKCTRL, action-info(S)) .
      (
         rec-msg(actionchooser, breakctrl, action(T?)) .
         snd-eval(BREAKCTRL, action(T)) .
         (
            rec-value(BREAKCTRL, break) .
            snd-msg(breakctrl, display, break(T)) .
            snd-msg(breakctrl, actionchooser, break)
         +  rec-value(BREAKCTRL, nobreak) .
            snd-msg(breakctrl, actionchooser, nobreak)
         )
      +  rec-msg(actionchooser, breakctrl, action-choose-list(S?)) .
         snd-eval(BREAKCTRL, action-choose-list(S)) .
```

```
              (
                 rec-value(BREAKCTRL, action-choose-list(S?)) .
                 snd-msg(breakctrl, actionchooser, action-choose-list(S))
              + rec-value(BREAKCTRL, break) .
                 snd-msg(breakctrl, display, break) .
                 snd-msg(breakctrl, actionchooser, break)
              )
          ) * delta
    endlet

    #undef linux
    #define QUOTE(X) #X
    #define XQUOTE(X) QUOTE(X)
    tool gui is {command = XQUOTE(wish-adapter -script LIBRARY/allgui.tcl) }
    tool kernel is {command = XQUOTE(perl-adapter -script LIBRARY/simkernel-adapter) }
    tool actionchooser(S : str) is {command = XQUOTE(wish-adapter -script LIBRARY/actionchooser.tcl
    tool function(S : str) is {command = XQUOTE(wish-adapter -script LIBRARY/function.tcl -script-ar
    tool display(S : str) is {command = XQUOTE(wish-adapter -script LIBRARY/display.tcl -script-args
    tool tracectrl(S : str) is {command = XQUOTE(wish-adapter -script LIBRARY/tracectrl.tcl -script-
    tool breakctrl(S : str) is {command = XQUOTE(wish-adapter -script LIBRARY/breakctrl.tcl -script-
    tool startprocess(S : str) is {command = XQUOTE(wish-adapter -script LIBRARY/process.tcl -script-a

    toolbus(PGUI, PKERNEL, PPROCESS, PACTIONCHOOSER, PFUNCTION, PDISPLAY, PTRACECTRL, PBREAKCTRL)
```

## B.2 Raku

```raku
#!/usr/bin/raku

use lib '/path_to_lib';
use BTCA;

my $toolpath = '/path_to_tools';

my %ch := Map.new(
   'gui-function', BTCA::Channel.new('gf'),
   'gui-startprocess', BTCA::Channel.new('gs'),
   'gui-tracectrl', BTCA::Channel.new('gt'),
   'gui-breakctrl', BTCA::Channel.new('gb'),
   'gui-display', BTCA::Channel.new('gd'),
   'gui-actionchooser', BTCA::Channel.new('ga'),
   'kernel-function', BTCA::Channel.new('kf'),
   'kernel-tracectrl', BTCA::Channel.new('kt'),
   'kernel-breakctrl', BTCA::Channel.new('kb'),
   'kernel-startprocess', BTCA::Channel.new('ks'),
   'kernel-actionchooser', BTCA::Channel.new('ka'),
   'kernel-display', BTCA::Channel.new('kd'),
   'kernel-gui', BTCA::Channel.new('kg'),
   'tracectrl-actionchooser', BTCA::Channel.new('ta'),
   'tracectrl-display', BTCA::Channel.new('td'),
   'breakctrl-actionchooser', BTCA::Channel.new('ba'),
   'breakctrl-display', BTCA::Channel.new('bd'),
   'actionchooser-tracectrl', BTCA::Channel.new('at'),
   'actionchooser-breakctrl', BTCA::Channel.new('ab'),
   'actionchooser-kernel', BTCA::Channel.new('ak'),
   'startprocess-kernel', BTCA::Channel.new('sk'),
   'display-kernel', BTCA::Channel.new('dk'),
   'function-kernel', BTCA::Channel.new('fk'),
);

sub Pgui() {
   my $windows;

   my $tool = BTCA::Tool.new('gui', '/usr/bin/wish', "$toolpath/allgui.tcl");
   $tool.Start();
   react {
      whenever $tool.Receive() {
         $_ ~~ %BTCA::expr{'rec-event'};
         $windows = $0;
         $tool.Send("snd-ack-event($windows)");
```

```perl
            $windows ~~ /window\((.*)\,\s*(.*)\,\s*(.*)\,\s*(.*)\,\s(.*)\,\s(.*)\)/;
            %ch{'gui-function'}.snd-msg("window($0)");
            %ch{'gui-startprocess'}.snd-msg("window($1)");
            %ch{'gui-tracectrl'}.snd-msg("window($2)");
            %ch{'gui-breakctrl'}.snd-msg("window($3)");
            %ch{'gui-display'}.snd-msg("window($4)");
            %ch{'gui-actionchooser'}.snd-msg("window($5)");
         }
      }
}

my $chinit = Channel.new;
sub Pkernel() {
   my $val;

   my $tool = BTCA::Tool.new('kernel', '/usr/bin/perl', "$toolpath/simkernel-adapter");
   $tool.Start();
   $tool.Send("snd-eval(get-action-info)");
   react {
      whenever $tool.Receive() {
         $_ ~~ %BTCA::expr{'rec-value'};
         my $m = $0;
         if $m ~~ /^action\-info\((.*)\)$/ {
            %ch{'kernel-tracectrl'}.snd-msg($m);
            %ch{'kernel-breakctrl'}.snd-msg($m);
            $tool.Send("snd-eval(get-process-list)");
         } elsif $m ~~ /^process\-list\((.*)\)$/ {
            %ch{'kernel-startprocess'}.snd-msg($m);
         } elsif $m ~~ /^action\-choose\-list\((.*)\)$/ {
            %ch{'kernel-actionchooser'}.snd-msg($m);
         } elsif $m ~~ /^halt\((.*)\)$/ {
            %ch{'kernel-actionchooser'}.snd-msg($m);
            %ch{'kernel-display'}.snd-msg($m);
         } elsif $m ~~ /^process\-status\((.*)\)$/ {
            %ch{'kernel-display'}.snd-msg($m);
         } elsif $m ~~ /^quit$/ {
            BTCA::Shutdown(0);
         }
      }
      whenever %ch{'startprocess-kernel'}.rec-msg() {
         if /^start\((.*)\,\s*(.*)\)$/ {
            %ch{'kernel-display'}.snd-msg("start($1)");
            $tool.Send("snd-do(start($0, $1))");
            $tool.Send("snd-eval(compute-choose-list)");
         } elsif /^reset$/ {
            $tool.Send("snd-do(myreset)");
            %ch{'kernel-display'}.snd-msg("reset");
            %ch{'kernel-actionchooser'}.snd-msg("reset");
            $tool.Send("snd-eval(compute-choose-list)");
         }
      }
      whenever %ch{'actionchooser-kernel'}.rec-msg() {
         if /^action\(info\((.*)\,\s*(.*)\,\s*
            $tool.Send("snd-do(action($0, $1,
            $tool.Send("snd-eval(compute-choose-list)");
         } elsif /^save\((.*)\)$/ {
            $tool.Send("snd-do($_)");
         } elsif /^goto\((.*)\)$/ {
            $tool.Send("snd-do(mygoto($0))");
            $tool.Send("snd-eval(compute-choose-list)");
         }
      }
      whenever %ch{'function-kernel'}.rec-msg() {
         if /^quit$/ {
            $tool.Send("snd-eval(quit)");
         } elsif /^process\-status$/ {
            $tool.Send("snd-eval(process-status)");
         }
      }
   }
}
```

```
sub Pprocess() {
   my $s = %ch{'gui-startprocess'}.rec-msg(:block);
   $s ~~ /^window\((.*)\)$/;
   my $winid = $0;
   my $tool = BTCA::Tool.new('process', '/usr/bin/wish', "$toolpath/process.tcl", "$winid");
   $tool.Start();
   react {
      whenever %ch{'kernel-startprocess'}.rec-msg() {
         if /^process\-list\((.*)\)$/ {
            $tool.Send("snd-do($_)");
         }
      }
      whenever $tool.Receive() {
         $_ ~~ %BTCA::expr{'rec-event'};
         my $m = $0;
         if $m ~~ /^start\((.*)\, (.*)\)$/ {
            %ch{'startprocess-kernel'}.snd-msg($m);
         } elsif $m ~~ /^reset$/ {
            %ch{'startprocess-kernel'}.snd-msg($m);
         }
         $tool.Send("snd-ack-event($m)");
      }
   }
}
sub Ptracectrl() {
   my $act;

   my $s = %ch{'gui-tracectrl'}.rec-msg(:block);
   $s ~~ /^window\((.*)\)$/;
   my $winid = $0;
   my $tool = BTCA::Tool.new('tracectrl', '/usr/bin/wish', "$toolpath/tracectrl.tcl", "$winid")
   $tool.Start();
   react {
      whenever %ch{'kernel-tracectrl'}.rec-msg() {
         my $s = $_;
         if $s ~~ /^action\-info\(.*\)$/ {
            $tool.Send("snd-do($s)");
         }
      }
      whenever %ch{'actionchooser-tracectrl'}.rec-msg() {
         my $s = $_;
         if $s ~~ /^action\((.*)\)$/ {
            $act = $0;
            $tool.Send("snd-eval($s)");
            $s = $tool.Receive(:block);
            if $s ~~ %BTCA::expr{'rec-value'} {
               my $m = $0;
               if $m ~~ /^trace$/ {
                  %ch{'tracectrl-display'}.snd-msg("trace($act)");
               }
            }
            %ch{'tracectrl-actionchooser'}.snd-msg("done");
         }
      }
   }
}
sub Pbreakctrl() {
   my $act;

   my $s = %ch{'gui-breakctrl'}.rec-msg(:block);
   $s ~~ /^window\((.*)\)$/;
   my $winid = $0;
   my $tool = BTCA::Tool.new('breakctrl', '/usr/bin/wish', "$toolpath/breakctrl.tcl", "$winid")
   $tool.Start();
   react {
      whenever %ch{'kernel-breakctrl'}.rec-msg() {
         my $s = $_;
         if $s ~~ /^action\-info\(.*\)$/ {
            $tool.Send("snd-do($s)");
         }
```

```
            }
            whenever %ch{'actionchooser-breakctrl'}.rec-msg() {
                my $s = $_;
                if $s ~~ /^action\((.*)\)$/ {
                    $act = $0;
                    $tool.Send("snd-eval($s)");
                } elsif $s ~~ /^action\-choose\-list\((.*)\)$/ {
                    $tool.Send("snd-eval($s)");
                }
                $s = $tool.Receive(:block);
                $s ~~ %BTCA::expr{'rec-value'};
                my $m = $0;
                if $m ~~ /^break$/ {
                    %ch{'breakctrl-display'}.snd-msg("break($act)");
                    %ch{'breakctrl-actionchooser'}.snd-msg("break");
                } else {
                    %ch{'breakctrl-actionchooser'}.snd-msg("nobreak");
                }
            }
        }
    }
    sub Pdisplay() {
        my $s = %ch{'gui-display'}.rec-msg(:block);
        $s ~~ /^window\((.*)\)$/;
        my $winid = $0;
        my $tool = BTCA::Tool.new('display', '/usr/bin/wish', "$toolpath/display.tcl", "$winid");
        $tool.Start();
        react {
            whenever %ch{'kernel-display'}.rec-msg() {
                my $s = $_;
                if $s ~~ /^process\-status\(.*\)$/ {
                    $tool.Send("snd-eval($s)");
                    $tool.Receive(:block); # rec value done
                } else {
                    $tool.Send("snd-do($s)");
                }
            }
            whenever %ch{'tracectrl-display'}.rec-msg() {
                $tool.Send("snd-eval($_)");
                $tool.Receive(:block); # rec value done
            }
            whenever %ch{'breakctrl-display'}.rec-msg() {
                if /^break\((.*)\)$/ {
                    $tool.Send("snd-do(break-action($0))");
                }
            }
        }
    }
    sub Pactionchooser() {
        my $Random = 0;
        my $acl;
        my $action;

        my $s = %ch{'gui-actionchooser'}.rec-msg(:block);
        $s ~~ /^window\((.*)\)$/;
        my $winid = $0;
        my $tool = BTCA::Tool.new('actionchooser', '/usr/bin/wish', "$toolpath/actionchooser.tcl", "
        $tool.Start();
        react {
            whenever %ch{'kernel-actionchooser'}.rec-msg() {
                my $s = $_;
                if $s ~~ /^action\-choose\-list\((.*)\)$/ {
                    $acl = $_;
                    if $Random {
                        %ch{'actionchooser-breakctrl'}.snd-msg($_);
                        my $b = %ch{'breakctrl-actionchooser'}.rec-msg(:block);
                        if $b ~~ /^break$/ {
                            $Random = 0;
                            $tool.Send("snd-do(random-off)");
                        }
```

```
                }
                $tool.Send("snd-do($acl)");
            } elsif /^halt\((.*)\)$/ {
                $tool.Send("snd-do(random-off)");
                $Random = 0;
                $tool.Send("snd-do(halt)");
            } elsif /^reset$/ {
                $tool.Send("snd-do(reset)");
            }
        }
        whenever $tool.Receive() {
            $_ ~~ %BTCA::expr{'rec-event'};
            my $m = $0;
            if $m ~~ /^random\((.*)\)$/ {
                $tool.Send("snd-ack-event($m)");
                if $0 eq "on" {
                    $Random = 1;
                } else {
                    $Random = 0;
                }
            } elsif $m ~~ /^save\((.*)\)$/ {
                $tool.Send("snd-ack-event($m)");
                %ch{'actionchooser-kernel'}.snd-msg($m);
            } elsif $m ~~ /^goto\((.*)\)$/ {
                $tool.Send("snd-ack-event($m)");
                %ch{'actionchooser-kernel'}.snd-msg($m);
            } elsif $m ~~ /^action\((.*)\)$/ {
                $tool.Send("snd-ack-event($m)");
                %ch{'actionchooser-kernel'}.snd-msg($m);
                if $Random {
                    %ch{'actionchooser-breakctrl'}.snd-msg($m);
                    $action = $m;
                    my $b = %ch{'breakctrl-actionchooser'}.rec-msg(:block);
                    if $b ~~ /^break$/ {
                        $Random = 0;
                        $tool.Send("snd-do(random-off)");
                    } elsif $b ~~ /^nobreak$/ {
                        %ch{'actionchooser-tracectrl'}.snd-msg($action);
                        %ch{'tracectrl-actionchooser'}.rec-msg(:block);
                    }
                } else {
                    %ch{'actionchooser-tracectrl'}.snd-msg($m);
                    %ch{'tracectrl-actionchooser'}.rec-msg(:block);
                }
            }
        }
    }
}

sub Pfunction() {
    my $s = %ch{'gui-function'}.rec-msg(:block);
    $s ~~ /^window\((.*)\)$/;
    my $winid = $0;
    my $tool = BTCA::Tool.new('function', '/usr/bin/wish', "$toolpath/function.tcl", "$winid");
    $tool.Start();
    react {
        whenever $tool.Receive() {
            $_ ~~ %BTCA::expr{'rec-event'};
            my $m = $0;
            $tool.Send("snd-ack-event($m)");
            if $m ~~ /^quit$/ {
                %ch{'function-kernel'}.snd-msg('quit');
            } elsif $m ~~ /^process\-status$/ {
                %ch{'function-kernel'}.snd-msg('process-status');
            }
        }
    }
}

BTCA::Run(&Pgui, &Pkernel, &Ptracectrl, &Pbreakctrl, &Pprocess, &Pdisplay, &Pactionchooser, &
```

### B.3 Go

```go
package main

import (
    "/path_to_lib/BTCA"
    "regexp"
)

var toolpath = "/path_to_tools"

var ch = map[string] chan string {
    "gf" : make(chan string),
    "gp" : make(chan string),
    "gt" : make(chan string),
    "gb" : make(chan string),
    "gd" : make(chan string),
    "ga" : make(chan string),
    "kf" : make(chan string),
    "kt" : make(chan string),
    "kb" : make(chan string),
    "kp" : make(chan string),
    "ka" : make(chan string),
    "kd" : make(chan string),
    "kg" : make(chan string),
    "ta" : make(chan string),
    "td" : make(chan string),
    "ba" : make(chan string),
    "bd" : make(chan string),
    "at" : make(chan string),
    "ab" : make(chan string),
    "ad" : make(chan string),
    "ak" : make(chan string),
    "pk" : make(chan string),
    "dk" : make(chan string),
    "fk" : make(chan string),
}

var expr = map[string] *regexp.Regexp {
    "windows" : regexp.MustCompile(`^window\((.*)\,\s*(.*)\,\s*(.*)\,\s*(.*)\,\s(.*)\,\s(.*)\`),
    "window" : regexp.MustCompile(`^window\((.*)\)$`),
    "start" : regexp.MustCompile(`^start\((.*)\,\s*(.*)\)$`),
    "action" : regexp.MustCompile(`^action\(info\((.*)\,\s*(.*)\,\s*(.*)\,\s*`),
    "action-single" : regexp.MustCompile(`^action\((.*)\)$`),
    "break" : regexp.MustCompile(`^break\((.*)\)$`),
    "random" : regexp.MustCompile(`^random\((.*)\)$`),
}

func Pgui() {
    tool := BTCA.NewTool("gui", "/usr/bin/wish", toolpath + "/allgui.tcl")
    tool.Start()
    run := 1
    for run == 1 {
        select {
        case a := <- tool.Receive:
            m := BTCA.Expr["rec-event"].FindStringSubmatch(a)
            tool.Send <- "snd-ack-event(" + a + ")"
            w := expr["windows"].FindStringSubmatch(m[1])
            if w != nil {
                ch["gf"] <- "window(" + w[1] + ")"
                ch["gp"] <- "window(" + w[2] + ")"
                ch["gt"] <- "window(" + w[3] + ")"
                ch["gb"] <- "window(" + w[4] + ")"
                ch["gd"] <- "window(" + w[5] + ")"
                ch["ga"] <- "window(" + w[6] + ")"
            }
        case a := <- ch["kg"]:
            if a == "quit" {
                run = 0
                tool.Kill()
```

```
            }
          }
        }
      }

    func Pkernel() {
        tool := BTCA.NewTool("kernel", "/usr/bin/perl", toolpath + "/simkernel-adapter")
        tool.Start()
        tool.Send <- "snd-eval(get-action-info)"
        run := 1
        for run == 1 {
          select {
          case a := <- tool.Receive:
              m := BTCA.Expr["rec-value"].FindStringSubmatch(a)
              if s, _ := regexp.MatchString("^action-info", m[1]); s {
                 ch["kt"] <- m[1]
                 ch["kb"] <- m[1]
                 tool.Send <- "snd-eval(get-process-list)"
              } else if s, _ := regexp.MatchString("^process-list", m[1]); s {
                 ch["kp"] <- m[1]
              } else if s, _ := regexp.MatchString("^action-choose-list", m[1]); s {
                 ch["ka"] <- m[1]
              } else if s, _ := regexp.MatchString("^halt", m[1]); s {
                 ch["ka"] <- m[1]
                 ch["kd"] <- m[1]
              } else if s, _ := regexp.MatchString("^process-status", m[1]); s {
                 ch["kd"] <- m[1]
              } else if s, _ := regexp.MatchString("^quit", m[1]); s {
                 BTCA.Shutdown()
              }
          case a := <- ch["pk"]:
              m := expr["start"].FindStringSubmatch(a)
              if m != nil {
                 ch["kd"] <- "start(" + m[2] + ")"
                 tool.Send <- "snd-do(" + a + ")"
                 tool.Send <- "snd-eval(compute-choose-list)"
              } else if s, _ := regexp.MatchString("^reset$", a); s {
                 tool.Send <- "snd-do(myreset)"
                 ch["kd"] <- "reset"
                 ch["ka"] <- "reset"
                 tool.Send <- "snd-eval(compute-choose-list)"
              }
          case a := <- ch["ak"]:
              m := expr["action"].FindStringSubmatch(a)
              if m != nil {
                 tool.Send <- "snd-do(action(" + m[1] + ", " + m[2] + ",
                 tool.Send <- "snd-eval(compute-choose-list)"
              } else if s, _ := regexp.MatchString("^save", a); s {
                 tool.Send <- "snd-do(" + a + ")"
              } else if s, _ := regexp.MatchString("^goto", a); s {
                 tool.Send <- "snd-do(my" + a + ")"
                 tool.Send <- "snd-eval(compute-choose-list)"
              }
          case a := <- ch["fk"]:
              if s, _ := regexp.MatchString("^quit$", a); s {
                 tool.Send <- "snd-eval(" + a + ")"
              } else if s, _ := regexp.MatchString("^process-status$", a); s {
                 tool.Send <- "snd-eval(" + a + ")"
              }
          }
        }
    }

    func Pprocess() {
        a := <- ch["gp"]
        m := expr["window"].FindStringSubmatch(a)
        tool := BTCA.NewTool("process", "/usr/bin/wish", toolpath + "/process.tcl", m[1])
        tool.Start()
        run := 1
        for run == 1 {
          select {
          case a := <- ch["kp"]:
```

```
            if s, _ := regexp.MatchString("^process-list", a); s {
               tool.Send <- "snd-do(" + a + ")"
            } else if a == "quit" {
               run = 0
               tool.Kill()
            }
         case a := <- tool.Receive:
            m := BTCA.Expr["rec-event"].FindStringSubmatch(a)
            if s, _ := regexp.MatchString("^start", m[1]); s {
               ch["pk"] <- m[1]
               tool.Send <- "snd-ack-event(" + m[1] + ")"
            } else if s, _ := regexp.MatchString("^reset$", m[1]); s {
               ch["pk"] <- m[1]
               tool.Send <- "snd-ack-event(" + m[1] + ")"
            }
         }
      }
   }

   func Ptracectrl() {
      var action string

      a := <- ch["gt"]
      m := expr["window"].FindStringSubmatch(a)
      tool := BTCA.NewTool("tracectrl", "/usr/bin/wish", toolpath + "/tracectrl.tcl", m[1])
      tool.Start()
      run := 1
      for run == 1 {
         select {
         case a := <- ch["kt"]:
            if s, _ := regexp.MatchString("^action-info", a); s {
               tool.Send <- "snd-do(" + a + ")"
            } else if a == "quit" {
               run = 0
               tool.Kill()
            }
         case a := <- ch["at"]:
            m := expr["action-single"].FindStringSubmatch(a)
            if m != nil {
               action = m[1]
               tool.Send <- "snd-eval(" + a + ")"
            }
         case a := <- tool.Receive:
            m := BTCA.Expr["rec-value"].FindStringSubmatch(a)
            if m != nil {
               if s, _ := regexp.MatchString("^trace$", m[1]); s {
                  ch["td"] <- "trace(" + action + ")"
               }
               ch["ta"] <- "done"
            }
         }
      }
   }

   func Pbreakctrl() {
      var action string

      a := <- ch["gb"]
      m := expr["window"].FindStringSubmatch(a)
      tool := BTCA.NewTool("breakctrl", "/usr/bin/wish", toolpath + "/breakctrl.tcl", m[1])
      tool.Start()
      run := 1
      for run == 1 {
         select {
         case a := <- ch["kb"]:
            if s, _ := regexp.MatchString("^action-info", a); s {
               tool.Send <- "snd-do(" + a + ")"
            } else if a == "quit" {
               run = 0
               tool.Kill()
            }
         case a := <- ch["ab"]:
```

```go
                m := expr["action-single"].FindStringSubmatch(a)
                if m != nil {
                    action = m[1]
                    tool.Send <- "snd-eval(" + a + ")"
                } else if s, _ := regexp.MatchString("^action-choose-list", a); s {
                    tool.Send <- "snd-eval(" + a + ")"
                }
            case a := <- tool.Receive:
                m := BTCA.Expr["rec-value"].FindStringSubmatch(a)
                if m != nil {
                    if s, _ := regexp.MatchString("^break$", m[1]); s {
                        ch["bd"] <- "break(" + action + ")"
                        ch["ba"] <- "break"
                    } else {
                        ch["ba"] <- "nobreak"
                    }
                }
            }
        }
    }

    func Pdisplay() {
        a := <- ch["gd"]
        m := expr["window"].FindStringSubmatch(a)
        tool := BTCA.NewTool("display", "/usr/bin/wish", toolpath + "/display.tcl", m[1])
        tool.Start()
        run := 1
        for run == 1 {
            select {
            case a := <- ch["kd"]:
                if s, _ := regexp.MatchString("^process-status", a); s {
                    tool.Send <- "snd-eval(" + a + ")"
                    <- tool.Receive
                } else if a == "quit" {
                    run = 0
                    tool.Kill()
                } else {
                    tool.Send <- "snd-do(" + a + ")"
                }
            case a := <- ch["td"]:
                // action(  )
                tool.Send <- "snd-eval(" + a + ")"
                <- tool.Receive
            case a := <- ch["bd"]:
                m := expr["break"].FindStringSubmatch(a)
                if m != nil {
                    tool.Send <- "snd-do(break-action(" + m[1] + "))"
                }
            }
        }
    }

    func Pactionchooser() {
        Random := 0
        var acl string
        var action string

        a := <- ch["ga"]
        m := expr["window"].FindStringSubmatch(a)
        tool := BTCA.NewTool("actionchooser", "/usr/bin/wish", toolpath + "/actionchooser.tcl", m[
        tool.Start()
        run := 1
        for run == 1 {
            select {
            case a := <- ch["ka"]:
                if s, _ := regexp.MatchString("^action-choose-list", a); s {
                    acl = a
                    if Random == 1 {
                        ch["ab"] <- a
                        b := <- ch["ba"]
                        if s, _ := regexp.MatchString("^break$", b); s {
                            Random = 0
```

```
                            tool.Send <- "snd-do(random-off)"
                        }
                    }
                    tool.Send <- "snd-do(" + acl + ")"
                } else if s, _ := regexp.MatchString("^halt$", a); s {
                    tool.Send <- "snd-do(random-off)"
                    Random = 0
                    tool.Send <- "snd-do(halt)"
                } else if s, _ := regexp.MatchString("^reset$", a); s {
                    tool.Send <- "snd-do(reset)"
                } else if a == "quit" {
                    run = 0
                    tool.Kill()
                }
            case a := <- tool.Receive:
                m := BTCA.Expr["rec-event"].FindStringSubmatch(a)
                if m != nil {
                    r := expr["random"].FindStringSubmatch(m[1])
                    if r != nil {
                        if r[1] == "on" {
                            Random = 1
                        } else {
                            Random = 0
                        }
                    } else if s, _ := regexp.MatchString("^save", m[1]); s {
                        tool.Send <- "snd-ack-event(" + m[1] + ")"
                        ch["ak"] <- m[1]
                    } else if s, _ := regexp.MatchString("^goto", m[1]); s {
                        tool.Send <- "snd-ack-event(" + m[1] + ")"
                        ch["ak"] <- m[1]
                    } else if s, _ := regexp.MatchString("^action", m[1]); s {
                        tool.Send <- "snd-ack-event(" + m[1] + ")"
                        ch["ak"] <- m[1]
                        if Random == 1 {
                            ch["ab"] <- m[1]
                            action = m[1]
                            b := <- ch["ba"]
                            if b == "break" {
                                Random = 0
                                tool.Send <- "snd-do(random-off)"
                            } else if b == "nobreak" {
                                ch["at"] <- action
                                <- ch["ta"]  //done
                            }
                        } else {
                            ch["at"] <- m[1]
                            <- ch["ta"]  //done
                        }
                    }
                }
            }
        }
    }

    func Pfunction() {
        a := <- ch["gf"]
        m := expr["window"].FindStringSubmatch(a)
        tool := BTCA.NewTool("function", "/usr/bin/wish", toolpath + "/function.tcl", m[1])
        tool.Start()
        run := 1
        for run == 1 {
            select {
            case a:= <- tool.Receive:
                m := BTCA.Expr["rec-event"].FindStringSubmatch(a)
                if m != nil {
                    if m[1] == "quit" {
                        ch["fk"] <- "quit"
                    } else if m[1] == "process-status" {
                        ch["fk"] <- "process-status"
                    }
                }
            case a := <- ch["kf"]:
```

```
            if a == "quit" {
                run = 0
                tool.Kill()
            }
        }
    }
}

func main() {

    BTCA.Run(Pgui, Pkernel, Ptracectrl, Pbreakctrl, Pprocess, Pdisplay, Pactionchooser, Pfunct
}
```